# Variations of the Near-Surface Electric field measured at Aragats during Geomagnetic Storms

A.Chilingarian

Yerevan Physics Institute, Alikhanyan Brothers 2, Yerevan, Armenia, AM0036

## Abstract

At least two mechanisms effectively transfer interplanetary magnetic field (IMF) disturbances into the atmosphere. First, the inflow of solar wind into the ionosphere at low latitudes significantly enhances the total vertical electron content, increasing atmospheric conductivity. Second, Forbush decreases (FD) reduce the cosmic ray flux by a few percent, lowering ionization levels at middle latitudes and decreasing conductivity. Changes in atmospheric conductivity affect the global electric circuit and atmospheric electric field (AEF). However, to study the response of AEF to geomagnetic storms (GMS), it is necessary to carefully monitor atmospheric conditions before and during storms, as meteorological influences can be much stronger than those of GMS. Charged clouds above detectors, lightning flashes, and abrupt weather changes significantly impact near-surface electric field (NSEF) variations, which serve as a proxy for AEF measured at the Earth's surface.

The facilities at Aragats station monitor all environmental parameters on a one-minute timescale. We analyze four GMS events described in previous studies, detailing the corresponding weather conditions to isolate the genuine influence of GMS on NSEF. The GMS of June 22, 2015, and September 8, 2017, occurred under fair-weather conditions, providing clear evidence of GMS influence on NSEF. These events were long-lasting, positive, and modest, ranging from 0.2 to 0.3 kV/m, and coincided with the depletion phase of FD. The sky was clear, no rain was detected, and lightning flashes from previous thunderstorms were more than 20 km from the station. The other two events did not meet favorable weather criteria, and their occurrence during GMS seemed incidental. We identify a feature that may indicate the solar (FD) origin of NSEF enhancement: a dip in the enhanced NSEF during the daytime.

## 1. Introduction

The solar wind profoundly influences Earth's magnetosphere, ionosphere, and global electric circuit (GEC) and modulates the flux of galactic cosmic rays reaching the atmosphere.

Geomagnetic storms (GMS), triggered by interactions between interplanetary coronal mass ejections (ICMEs) and the magnetosphere, induce electric currents in Earth's environment and enhance radiation belts.

To establish causal relationships between GMS and GEC variations, we analyze key parameters: the interplanetary magnetic field (IMF) and its Bz component (embedded within the ICME), the horizontal component of the geomagnetic field (GMF) Bx, the vertical component of the atmospheric electric field (Ez), and the flux of secondary cosmic rays arriving at Earth's surface. The IMF is measured by magnetometers at Lagrangian points and GOES satellites, Ez by near-surface electric field sensors, Bx by ground-based magnetometers, and cosmic ray flux by particle detectors and spectrometers.

This study uses IMF data from the WIND satellite and Bx, Ez, and particle flux data from the Aragats Research Station of the Yerevan Physics Institute [1]. To accurately assess the influence of GMS on Ez, we focus on data collected under "fair weather" conditions, isolating it from the impact of local weather-related fluctuations in Ez and particle fluxes. Since electric field measurements are significantly affected by local charge distributions and inhomogeneities of electrical conductivity, separating these effects from the more subtle influences possibly induced by GMS is crucial.

The overall intensity of geomagnetic storms is quantified by the disturbance storm time index (DST), a Bx average from four near-equatorial observatories that represent the ring current surrounding Earth. The planetary Kp index estimates global geomagnetic activity levels caused by solar wind and IMF variations. Kp is compiled from measurements taken every three hours at multiple geomagnetic observatories around the world. Each observatory assigns a local K index, which is then averaged to obtain the planetary Kp index.

The influence of GMS on Ez variations was first reported in [2]; authors documented substantial (~100–300 V/m) decreases in Ez at the Swider Observatory in Poland, concurrent with substorm onset in auroral latitudes.

The effects of the geomagnetic storms of November 8 and 10, 2004, were studied at Kamchatka's Paratunka observatory (52.9° N, 158.25° E), simultaneously observing variations in the atmospheric electric field (AEF) strength and meteorological parameters in the near Earth's atmosphere [3]. An enhancement of the strength of power spectra of the electric field was detected at the commencement of the geomagnetic storm of November 10. The authors explained this effect due to the action of cosmic rays on global electric circuit currents by changing the ionization of the atmosphere in different parts of the circuit, solar cosmic rays (SCR) at heights of ~50 km and galactic cosmic rays (GCRs) at heights of ~15–20 km.

The Borok Geophysical Observatory of the Institute of Physics of the Earth of the RAS (58°04′ N, 38°14′ E) detected atmospheric electric field increase during GMS on March 31, 2001, and May 15, 2005 (Fig. 1 of [4]) correlated with the storm severity.

A study at Mohe Station in China [5], based on 15 GMS events, found that Ez decreased by 100–600 V/m when Bz shifted southward. The authors claim that variations in cosmic ray intensity during GMS could impact lower-atmosphere conductivity.

Recently, [6] reported anomalous Ez variations during the GMS on April 24, 2023, using data from eight mid- and low-latitude observation stations in China. Their analysis revealed Ez values fluctuating from 19 to 370 V/m (with daily means of 10–260 V/m) relative to the global average of Bx near the magnetic equator (DST index), with delays ranging from 0 to 5.3 hours. Only one station observed Ez anomalies synchronized with variations; five stations reported increases in Ez, and three reported decreases. Thus, the data is rare and controversial due to the substantial "noise" overlaying the relatively small expected effects of atmospheric conductivity. To overcome these difficulties, we investigate correlations between all available types of measurements—IMF, GMF, Ez, Kp, lightning occurrences during and before GMS, meteorological parameters, clouds above the detectors, and cosmic ray flux. All these parameters (except IMF and KP) are taken at the Aragats Research Station during observations of the most intense GMS of the 24th and 25th solar cycles.

Temporary magnetic structures can form during nonlinear interactions between ICMEs and the magnetosphere. These structures confine GCRs, significantly reducing the flux of secondary particles reaching surface detectors—an effect known as the Forbush decrease (FD). FD can lead to a large depletion of secondary CR flux, which depends on geomagnetic cutoff rigidity (Rc). This rigidity is high at low latitudes and low at high and middle latitudes. At Aragats, the value is 7.1 GV.

Particle detectors at Aragats accurately measure secondary muons resulting from interactions of GCR and SCR. These secondary fluxes, in turn, influence the atmospheric-ionospheric currents, which cause variations in the near-surface electric field (NSEF).

While our previous studies have detailed solar particle events of the 25th solar activity cycle [7,8]. This study concentrates on Ez variations caused by FD (GMS).

## 2. Instrumentation

Below is a brief description of the instruments used at the Aragats Observatory to measure the atmospheric electric field (AEF), geomagnetic field (GMF), meteorological parameters, lightning location, and particle fluxes.

The near-surface electric field (NSEF, a term used interchangeably with the vertical atmospheric electric field $EzE\_zEz$) is continuously monitored by a network of commercially available field mills (EFM-100) [9]. Three of these are installed at the Aragats station, one at the Nor Amberd station, located 12.8 km from Aragats, one in Burakan village, 15 km from Aragats, and one at the Yerevan station, 39.1 km from Aragats.

The electric field polarity follows the standard meteorological convention (AEF convention [10]), where a positive reading indicates an electric field directed downward (from the atmosphere to the ground), which is typical in fair-weather conditions. A negative reading suggests an upward electric field (from the ground to the atmosphere), common during stormy conditions or near charged clouds.

The GEC current is controlled by the ionospheric potential (~250 kV), not by local conductivity variations. The electric field adjusts accordingly since the global circuit tries to maintain a steady current. Thus, when atmospheric conductivity increases, charges redistribute more efficiently, reducing the near-surface electric field. This results in lower EFM-100 readings. Conversely, if the downward current density in the atmospheric column decreases due to lower ionization levels, atmospheric conductivity decreases, leading to a stronger electric field near the surface. In the AEF convention, NSEF becomes more positive with reduced conductivity. The existence of two conventions for electric field polarity can sometimes cause confusion. Suppose an electric field sensor follows the physics convention rather than the meteorological one. In that case, its display will be inverted, meaning that positive and negative values will appear opposite to those in the AEF convention.

The distances between the three field mills at Aragats are 80 m, 270 m, and 290 m (see Fig.1). The sensitivity distance of EFM-100 for the lightning location is 33 km, and the response time of the instrument is 100 ms. The electrostatic field changes are recorded at a sampling interval of 50 ms. Data on the NSEF field starting from 2010 (1s time series) are available via the Advanced Data Extraction Infrastructure (ADEI) [11].

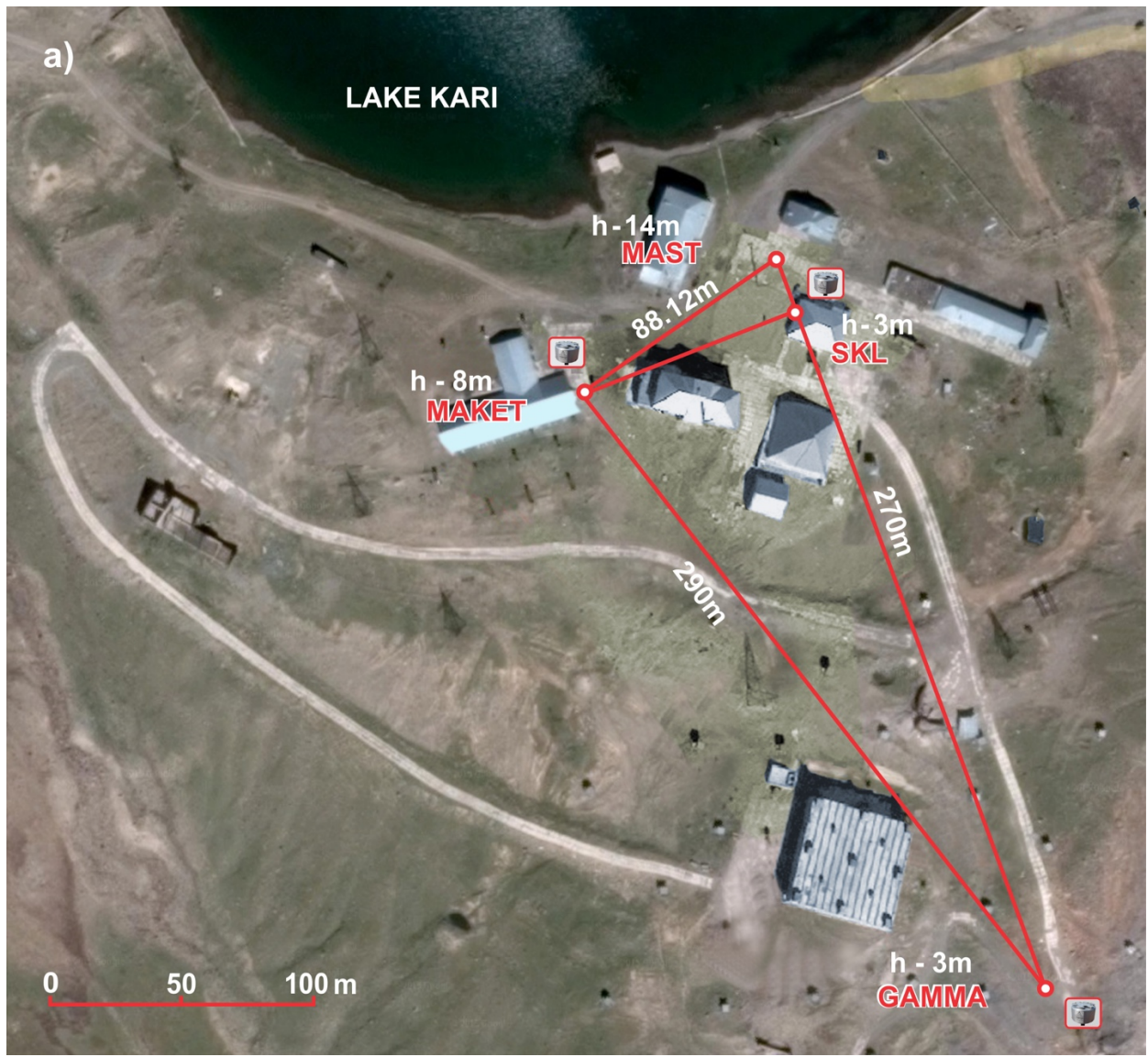

**Figure 1.** Network of electric mills EFM 100 on Aragats station (3200 m).

The lightning activity from 30 km to 480 km is monitored by Boltek's Storm Tracker [12]. Storm tracker defines four types of lightning types (CG-, CG+ cloud-to-ground negative and positive, IC-, IC+ intracloud positive and negative) in radii up to 480 km around the location of its antenna. By examining the time-slices of the lightning activity, we determine from which direction the storm is coming, and, finally, by putting on the map all lightning occurrences, we can see if the storm's active zone goes above the stations or misses them.

The LEMI-417M digital magnetotelluric station, produced by Lviv Space Research Institute [13], has operated on Aragats since September 2010. It provides a one-minute time series of the three components of the geomagnetic field. The magnetometer is based on the flux-gate sensor, and all three components are implemented in the same thermostable housing.

Davis Instruments' Vantage Pro2 Plus automatic weather stations measure meteorological conditions. These stations include a rain collector, temperature and humidity sensors, an anemometer, a solar radiation sensor, and a UV sensor. They are in Aragats (two units), Nor Amberd, Burakan, and Yerevan. One-minute time series are available from 2010.

The ALL-SKY CAM panoramic cameras Moonglow Technologies produces 24/7 monitor the clouds above Aragats station. A one-minute time series of camera shots is accessible from the ADEI (starting in 2012). A circular fisheye system provides a 190° hemispherical field of view. The image sensor is a Color 1/3" Sony Super HAD CCD II with an effective pixel number across FoV of 546 × 457, with an automatic exposure time (from $10^{-5}$s to 4s).

The Aragats Neutron Monitor (ArNM, data from 2003), type 18HM64, and the SEVAN and muon detectors (data from 2008) are located in the MAKET experimental hall. A network of seven spectrometers (based on NaI crystals of 12 × 12 × 28 cm size) and STAND3 and CUBE stacked scintillation detectors are located in the SKL experimental hall. Aragats spectrometers measure the energy spectrum of secondary CR [14]. During FD, by subtracting the depleted particle flux from the fair-weather one measured before FD, we gained insight into the energies of primary GCR that were effectively trapped by the disturbances in near-Earth plasma during the interaction of IMF and GMF. Consequently, by simulating GCR protons traversing the terrestrial atmosphere, we can obtain the most probable energies of primary particles for each species of secondary CR [15].

### 3. Physical model of atmosphere ionization during solar events

Two mechanisms during geomagnetic storms (GMS) can influence the near-surface electric field (NSEF):

1. The coupling of the southward interplanetary magnetic field with the geomagnetic field allows enhanced penetration of solar wind particles into the polar ionosphere. This influx of energetic particles dramatically increases ionization in the upper atmosphere, particularly at high latitudes, leading to enhanced conductivity and a reduction of the NSEF due to the lowering of atmospheric column resistance.
2. The Forbush Decrease results from the shielding of galactic cosmic rays by interplanetary ejecta during a GMS. As GCRs are the dominant ionization source in the middle and lower atmosphere, their temporary reduction leads to a decrease in ion production, which increases atmospheric resistance and could enhance NSEF at low to mid-latitudes.

FDs typically reduce GCR flux by 3–10%, depending on geomagnetic cutoff rigidity and the storm's intensity [16]. According to atmospheric ionization models [17], this results in a **5–7%** decrease in ion pair production rates at altitudes of 10–20 km. However, the impact of such modest reductions on electric field strength is not directly proportional. It depends strongly on local atmospheric conditions**,** vertical conductivity profiles, and the structure of the global electric circuit.

Theoretical models (e.g., [18]) have demonstrated that small changes in conductivity can modulate vertical current flow in the global electric circuit. However, the relationship between reduced ionization and measurable enhancements in NSEF is highly nonlinear and sensitive to cloud microphysics, temperature inversions, and charge redistribution. Localized enhancements in NSEF during FDs can be amplified through secondary effects, rather than arising solely from the decrease in ionization.

In contrast, the ionization caused by solar wind particles during strong GMS events can be much larger than the ionization change from FDs, particularly in the polar ionosphere. This leads to significant conductivity enhancements and often results in substantial negative changes in NSEF at high latitudes [19].

Therefore, while both mechanisms—GCR suppression and solar wind particle influx—modulate atmospheric ionization, they act at different latitudes and altitudes, with **different magnitudes**. GCR reduction produces a minor, diffuse effect **in** the middle atmosphere, while solar particles induce intense, localized ionization at high altitudes.

Observed NSEF variations at middle latitudes during FDs are thus expected to be small and positive, but often modulated or dominated by local meteorological conditions, such as:

- the presence of stratified or charged clouds,
- lightning activity or precipitation,
- and rapid changes in humidity or temperature structure.

Consequently, identifying and quantifying the contribution of GMS-related drivers to NSEF variations requires high-resolution, multi-parameter observations, such as those provided at Aragats station. In the following sections, we analyze four GMS events using one-minute resolution measurements of cosmic ray flux, NSEF, and atmospheric conditions.

As a clarification, the polarity of the electric field measurements follows the standard Atmospheric Electric Field (AEF) convention.

4. **Variations of the Near-Surface Electric field during GMS: 4 showcases**

On June 22, 2015, Earth experienced a significant geomagnetic storm, one of the most intense of Solar Cycle 24. This geomagnetic storm was initiated by a series of halo coronal mass ejections originating from Active Region AR 12371 between June 18 and June 25, 2015. The ICMEs impacted Earth's magnetosphere, leading to a G4-class geomagnetic storm on June 22–23, 2015. The ICME associated with this storm had a speed of 700–800 km/s upon impact. The sudden storm commencement (SSC), marking the onset of the storm, was recorded on June 22, 2015, at 18:37 UTC. The decaying phase of the storm began shortly after the SSC, when Earth's magnetic field responded to the incoming ICME, and ended at 23:30 on the same day, lasting approximately 5 hours. However, the reduced flux of the ionizing particles continued for an additional 18 hours due to the arrival of subsequent ICMEs. This reduction in cosmic ray intensity observed by ground-based neutron monitors was notable, reaching approximately 6% at middle latitudes [20]. Scintillators registering muon flux showed a 4% reduction compared to neutron monitors, indicating a different response of secondary cosmic ray components to geomagnetic disturbances. Figure 2 illustrates the disturbances (black) alongside the muon count rate (blue). The inset reveals the clear sky above Aragats. As expected, the enhancement of NSEF was positive and relatively modest, around 0.2 kV/m, compared with the fair-weather value of 0.13 ± 0.2 kV/m. It lasted from 08:00 to 18:00 UTC on June 23, with a break during the daytime from 10:00 to 15:00. The flux of secondary muons (a proxy of GCR entering the atmosphere) was minimal during NSEF enhancement. No rain or clouds were detected above the station. The outside temperature fluctuated from 10°C to 13°C, atmospheric pressure remained stable at approximately 694 mb, and relative humidity rose from 50% to 80%. A remote lightning flash, located 25 km from the station, was detected on June 22 at 12:00, about seven hours before the ICME arrival. No clouds were detected above the detectors during NSEF enhancements, as illustrated in the inset to Fig. 2.

Data from Swarm satellites show a significant daytime increase in the total vertical electron content during the storm's initial phase, around 19:00–21:00 UTC on June 22, 2015, and at the end of its main phase [21].

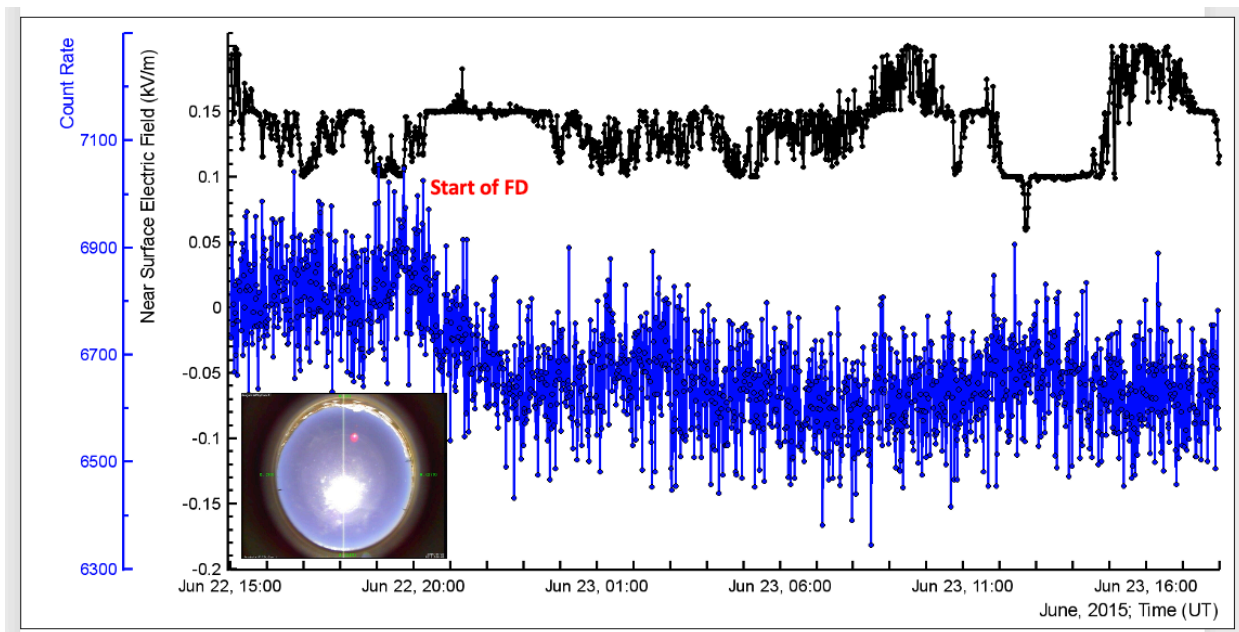

**Figure 2. Disturbances of NSEF (black) and count rate measured by a 3 cm thick 1 m$^2$ area plastic scintillator (blue) on June 22, 2015. In the inset, we show the sky above the station.**

On September 6, 2017, a powerful X9.3 solar flare erupted from Active Region 12673 at 12:02 UTC. The flare was accompanied by a fast and massive full-halo ICME directed toward Earth, with an estimated speed of approximately 1,500 km/s. The SSC, indicating the arrival of the ICME's shock front, was observed at 23:43 UTC on September 7, 2017. Figure 3 shows the time series of the same variables as in Figure 2. The inset displays the blue sky above Aragats. The decaying phase 0f FD was long, approximately 15 hours; the positive enhancement NSEF coincides with decreased muon flux.

NSEF disturbances occurred between 04:00 and 16:00 on September 8, in line with the ongoing FD with an amplitude of ≈5%. The NSEF enhancement was moderate, ranging from 0.25 to 0.3 kV/m, compared to the mean value measured before the GMS, which was 0.13 ± 0.02 kV/m. A deepening of NSEF was again observed from 10:00 to 14:00 UTC during the daytime.

At 10:53 on September 6, a lightning flash was detected 20 km from the station; the air above the detectors was clear. Over 24 hours, the outside temperature increased from 2°C to 8°C, atmospheric pressure changed from 694 to 696 mb, and relative humidity rose from 50% to 60%. No rain was detected.

In [22], the same event was analyzed using data from the JCI 131F sensor (Azores, AZO station, 39.09°N, 28.03°W, altitude 31 m, Rc = 6.6) and the EFM 100 electric field mill (Studenec, STU station, 50.26°N, 12.52°E, altitude 712 m, Rc = 3.31). Comparisons were made with neutron flux measured by OULU's NM (65.05°N, 25.47°E, altitude 15 m, Rc = 0.8).

The authors concluded that NSEF at both stations followed the FD, decreasing by 0.1 kV/m at the AZO station and 0.015 kV/m at the STU station. The depletion started at 00:00 UTC on September 8, reached a minimum at 06:00, and then significantly increased. However, the depletion was relatively small, and the decrease was not highly significant due to the large fluctuations in NSEF observed in Figure 4 (Li et al., 2023).

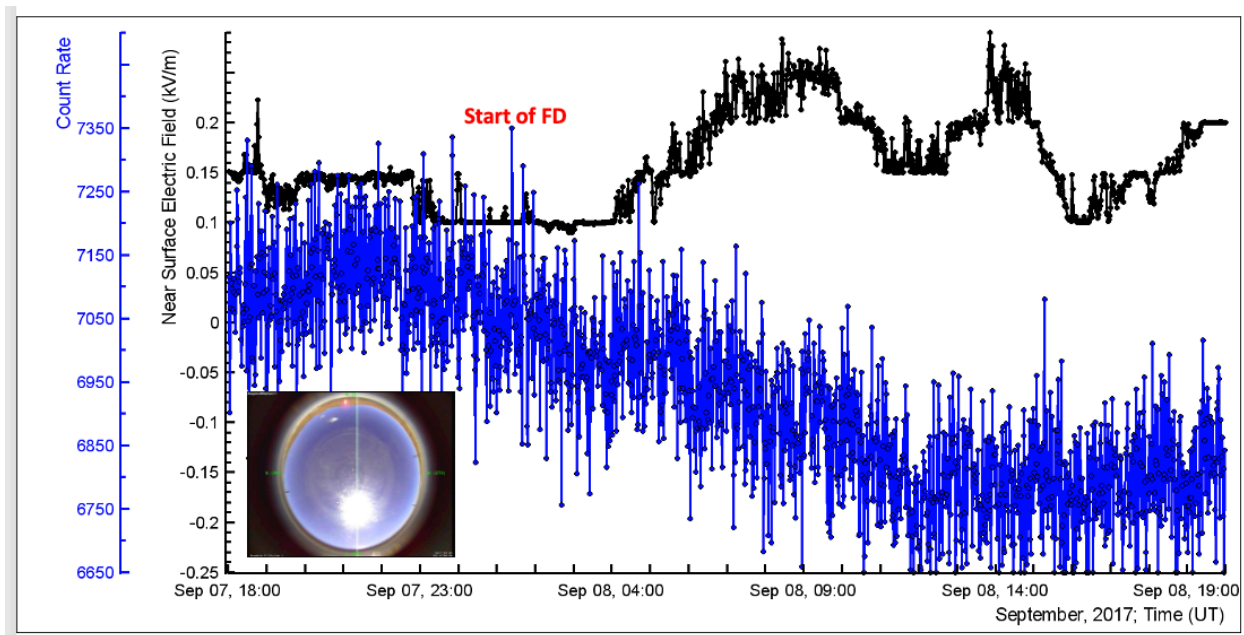

**Figure 3. Disturbances of NSEF (black) and the count rate of a 3 cm thick, 1 m² scintillator (blue) on September 6, 2017; inset – a photo of the sky above the station.**

On November 3–5, 2021, a large GMS unleashed auroras as far south as New Mexico (39°N). SOHO coronagraphs captured the storm cloud leaving the Sun on November 2, following and overtaking a slower-moving solar flare (M1.7) in the magnetic canopy of sunspot AR2891.

In Figure 4, we show the disturbances of NSEF (black) and the count rate of the STAND3 particle detector (blue). The "cannibal" ICME approached and passed the satellites at the L1 point on November 3; the SSC was recorded at 19:57 UTC. The depletion of NSEF coincided with the particle flux decrease during the declining phase of the FD. The NSEF deepened to negative values at approximately 09:00 and returned to positive values around 10:00, reaching a minimum of roughly -0.6 kV/m at 09:20 and 09:50. Before the GMS, during fair weather, the NSEF was 0.11 ± 0.03 kV/m. During NSEF depletion, no precipitation was recorded; however, the clouds above the station appeared (see inset to Figure 4). As multiyear observations show, even white (non-thunderstorm) clouds can induce NSEF with magnitudes reaching 1 kV/m. The last lightning flash, preceded by 2.5 hours of light rain, was detected 20 km from the Aragats station at noon on November 3. The outside temperature fluctuated between -1.5°C and 3.5°C, atmospheric pressure between 695 and 698 mb, and relative humidity between 72% and 95%.

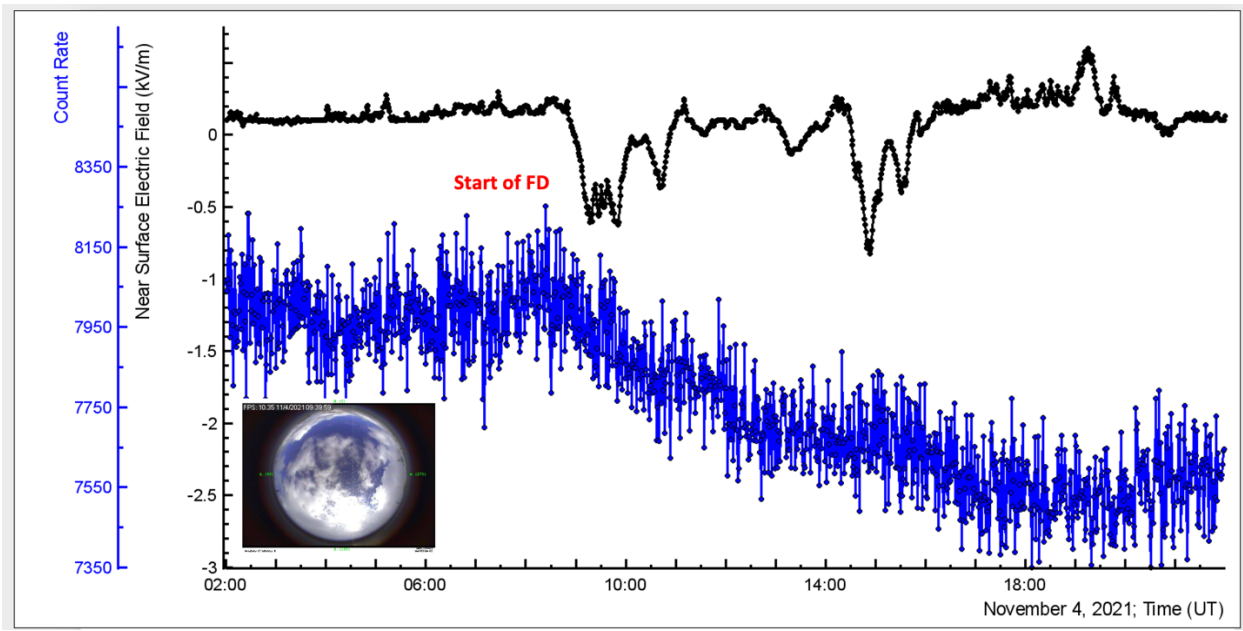

**Figure 4. Disturbances of NSEF (black), and the count rate of a 3 cm thick, 1 m² scintillator (blue) on November 4, 2021; inset – a photo of the sky above the station.**

On April 21, 2023, at 18:12 UTC, the Sun emitted an M1.7-class solar flare from AR13283. This flare was accompanied by a filament eruption and a full-halo ICME, indicating that the ejected material was directed toward Earth. On April 23, 2023, at 17:37 UTC, the ICME impacted Earth's magnetosphere, initiating a G4 GMS. The SSC was observed at the same time.

April on Aragats is marked by intense thunderstorms, during which the NSEF strength fluctuates between -30 and +20 kV/m. This variation is over an order of magnitude greater than expected due to FD; refer to Sections 3 and Figures 2 and 4 (black curves). In Figure 5, we observe two significant fluctuations in NSEF that coincide with bursts of particle flux. This phenomenon is due to operation of powerful electron accelerator in thunderclouds and is known as thunderstorm ground enhancement (TGE) [23,24].

TGE involves large fluxes of electrons, gamma rays, and neutrons generated by avalanches triggered by runaway electrons [25] when the atmospheric electric field exceeds the critical strength. The thundercloud during TGE typically hovers low above the Earth's surface, allowing particles to reach the ground before decaying through interactions with dense air. As shown in the inset of Figure 5, the cloud (fog) was positioned directly above the station.

Therefore, the large outbursts of NSEF depicted in Figure 5 (black curve) are attributed to thunderstorms, not FD. This event illustrates the importance of measuring as many environmental parameters as possible to establish a causal relationship between GMS/FD and NSEF.

The April 2023 event has also been reported in [6], which observed a significant decrease in the NSEF at some stations and a noticeable increase at others in middle latitudes in China on 24 April. This time, they utilized a DST time series (since no FD was registered) for comparison with NSEF data from Ganjingzi Station (GJZ, 39.0° N, 121.7° E) and Jinpuxin Station (JPX, 38.6° N, 121.5° E), which are 47.7 km apart. They noted a decrease of 0.2 kV/m at 4:00 on 24 April, followed by an increase that reached 0.4 kV/m at 5:00. The NSEF measured at the other station decreased by 0.05 kV/m, hitting a minimum of 0.1 kV/m at approximately 9:00. Once again, there is a discrepancy in the sign of NSEF fluctuations at nearby stations, and given the absence of FD, the most likely cause of the small fluctuations is random variations.

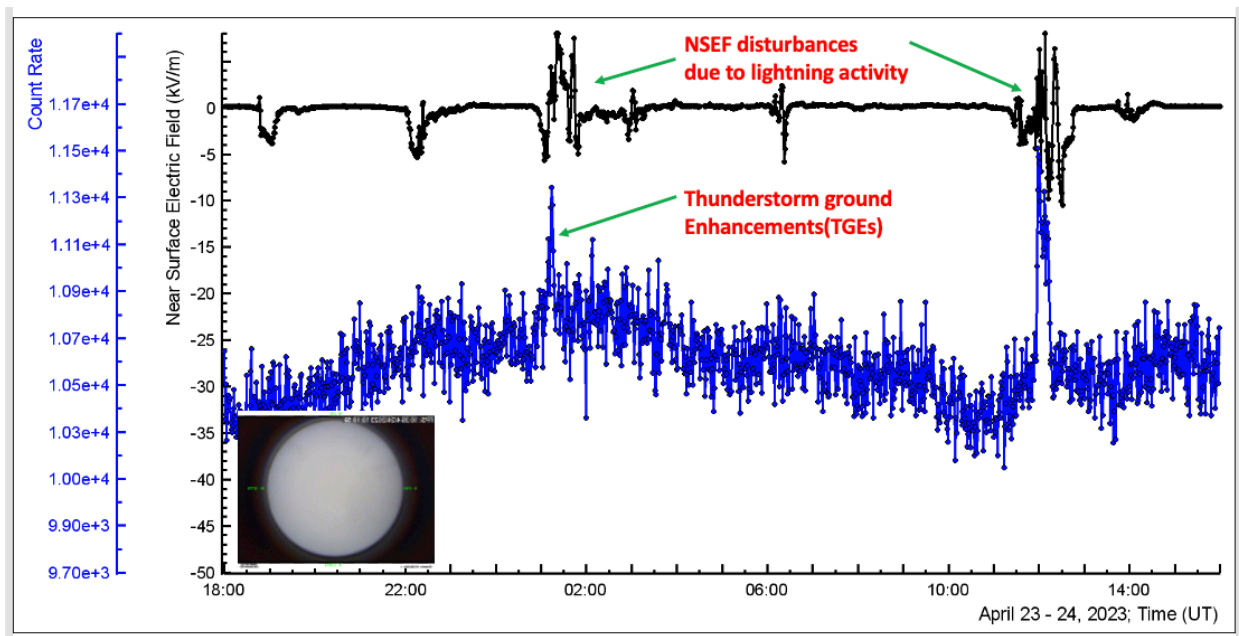

Figure 5. Disturbances of NSEF (black), and count rate of the STAND3 detector (blue) on **April 21, 2023**; the inset shows cloud sitting on the station.

## 5. Detailed analysis of the relation GMS-FD-NSEF

Among the four events presented in the previous section, only the first two demonstrate a small positive enhancement of NSEF during minimal values of secondary muon flux. This behavior aligns with the physical model described in Section 3. GMS unleashed FD, which, in turn, reduces the ionization and conductivity of the atmosphere, thus enhancing the potential between the ionosphere and Earth. Despite the strong GMS, the NSEF variations on November 4, 2021, and April 23-24, 2023, did not show the correct direction of the NSEF variation and were accompanied by fluctuating atmospheric conditions. These conditions, especially on April 23-24, influence NSEF and muon flux much more dramatically than any effect GMS could achieve. Thus, we continue the detailed analysis of the two selected events, discarding the others.

In Figure 6, we present the disturbances of the IMF (panels a and b), the geomagnetic field (panel c), the Kp index (panel d), the NSEF (panel e), and the muon count rate (panel f). In the left part of Figure 6, we present the event that took place on June 22-23, 2015. B and Bz were measured by the WIND spacecraft's magnetometer [26]. The B value reached peaks of around 40 nT during the ICME's arrival. The Bz component fluctuated strongly, reaching negative values of approximately -40 nT, which facilitated reconnection and allowed energy to flow into the magnetosphere. The southward orientation, opposite Earth's magnetic field, is particularly effective in coupling solar wind energy into the magnetosphere, leading to intensified geomagnetic activity. Kp values are measured by the German Research Center for Geosciences (Potsdam) as the prolonged GMS reached the level of a severe storm at 18:00 UTC. The geomagnetic field's X component, measured by the Aragats magnetometer [27], turns negative during the FD. ICME compresses the geomagnetic field and creates traps that prevent low-energy GCR from entering the atmosphere.

In the right panel, we display the event that took place on September 7-8, 2017. At approximately 20:00 UTC on September 7, the Bz component shifted abruptly to a southward (negative) orientation, reaching a minimum of -30 nT. The B value simultaneously reached 30 nT, sustaining G4-class geomagnetic storms. The early extrema of B and Bz suggest that strong magnetic field fluctuations occurred as part of an interaction with a dense region in the solar wind preceding the main ICME shock. This led to an unusual situation where significant geomagnetic effects were observed before the SSC. The FD started at approximately 20:00 on September 7. The Bx component began to diminish with the arrival of the ICME, reaching a first minimum at 07:00 and a second, deeper minimum at 13:00 on September 8. Correspondingly, the muon flux diminished.

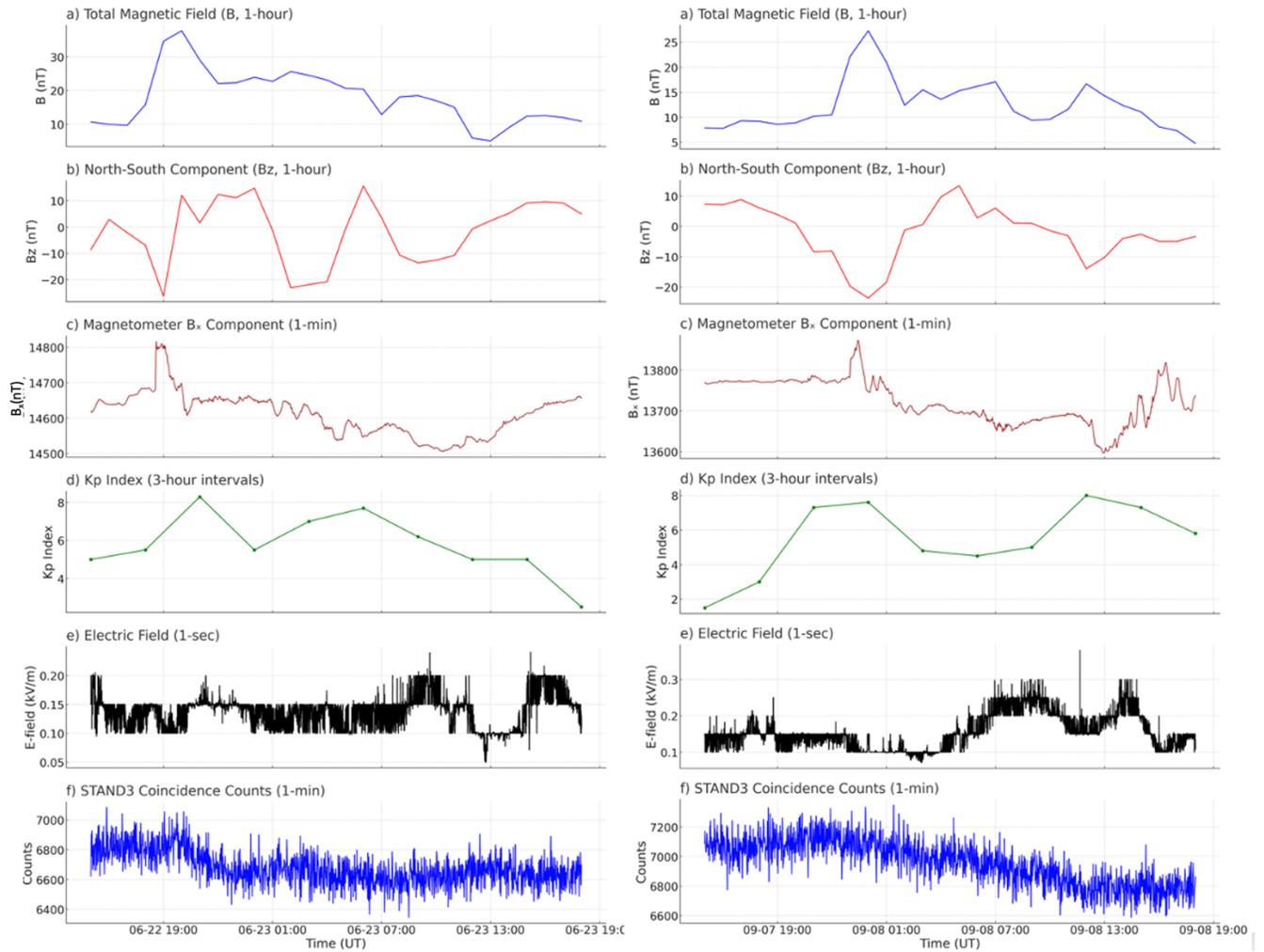

**Figure 6.** Time series of parameters related to strong GMSs on 22-23 June 2015 (left) and 7-8 September 2017 (right). a) and b) - IMF parameters by the WIND spacecraft's magnetometer [26]; c) X component of the geomagnetic field measured by the Aragats magnetometer [27]; d) - Kp values measured by the German Research Center for Geosciences (Potsdam); e) and f) – NSEF measured by EFM-100 electric mill, and particle flux measured by a 3 cm thick 1 m$^2$ area plastic scintillator on Aragats [27].

**Discussion and Conclusions**

The inflow of solar wind into the ionosphere at low latitudes significantly enhances the total vertical electron content, increasing atmospheric conductivity and decreasing the near-surface electric field at near-polar latitudes. The second mechanism involves a decrease in cosmic ray flux due to a Forbush decrease, reducing atmospheric ionization at middle latitudes. This reduction in ionization increases the potential drop between the ionosphere and the Earth's surface, potentially strengthening the near-surface electric field.

This study examines whether intense geomagnetic storms, triggered by interplanetary shocks and magnetospheric interactions, impact the atmospheric electric field observed at middle latitudes. Previous research has yielded mixed results that do not always align with theoretical expectations (see the description of events in Section 4). The expected effect of a Forbush decrease is small, as the flux of GCR diminishes by only a few percent. Meteorological factors often overshadow this effect. Therefore, selecting solar events where the atmospheric-ionospheric influence can be distinguished from atmospheric noise is critical.

Variations in the near-surface electric field are linked to current meteorological conditions and preexisting atmospheric conditions. At the Aragats research station, located 3,200 meters above sea level, we continuously monitor multiple environmental parameters to ensure that the selected geomagnetic storm events occur under fair-weather conditions. This provides a reliable reference for assessing their impact on the near-surface electric field.

The geomagnetic storms of 2015 and 2017, which occurred under fair-weather conditions, provide reliable evidence of FD influence on the near-surface electric field. The enhancement of NSEF occurred during the decaying phase of a Forbush decrease, hours after the interplanetary magnetic field reached extreme values. The observed electric field enhancements were positive, of 0.2 to 0.3 kV/m, compared to a stable fair-weather value of 0.13 kV/m. The X-component of the geomagnetic field decreased (c panel of Fig. 6), indicating strong pressure of the interplanetary magnetic field that disturbs the magnetosphere and prevents low-energy galactic cosmic rays from entering the atmosphere. High scalar interplanetary magnetic field values (a panel of Fig.6) support a Forbush decrease. The sky was clear during these events, with no rainfall recorded, and previous thunderstorms had been more than 20 kilometers away from the station.

Based on these observations, the positive enhancement of the near-surface electric field during the geomagnetic storms of June 22, 2015, and September 7, 2017, can be attributed to solar-driven variations in tropospheric ionization caused by the FD effect. From these two events, we can outline a feature indicating the solar origin of the NSEF enhancement: a dip in the enhanced NSEF during the daytime. This dip can be attributed to ultraviolet solar radiation compensating for the reduced air ionization caused by the FD effect. Additionally, the daytime dip in NSEF may also result from changes in heating and humidity rather than direct ultraviolet compensation for FD-related ionization loss.

In contrast, the geomagnetic storms of 2021 and 2023 differed significantly from the first two cases, as fair-weather conditions could not be confirmed. The near-surface electric field was negative, the duration of enhancements was very short, and the observed values were excessively high, ranging from 0.6 to 4 kV/m. During these events, clouds and fog indicated strong meteorological interference. Previous studies have shown that a single cloud can induce a strong local electric field even in winter, which may sufficiently trigger a thundercloud ground enhancement. Furthermore, in 2021, no Forbush decrease was observed, suggesting no direct solar influence on the near-surface electric field.

These findings suggest that geomagnetic storms on June 22, 2015, and September 6, 2017, influence the near-surface electric field through modifications in atmospheric conductivity. However, other environmental factors, including meteorological conditions, dominate the short-duration and high-amplitude electric field variations observed on November 4, 2021, and April 21, 2023. Future studies should further investigate the statistical significance of these effects and evaluate the relative contributions of solar and atmospheric influences.

The data supporting this study are available in both numerical and graphical formats through the multivariate visualization software platform ADEI, which is hosted on the Cosmic Ray Division (CRD) webpage of the Yerevan Physics Institute [28].

**ACKNOWLEDGMENTS**

We sincerely thank the staff of the Aragats Space Environmental Center for their seamless operation of the experimental facilities on Mount Aragats. The ADEI platform, developed by S. Chilingaryan, was very useful in finding appropriate data across multiyear and multiparameter observations. The author thanks the anonymous referee and the support of the Science Committee of the Republic of Armenia for Research Project No. 21AG-1C012

**Glossary**

- **Advanced Data Extraction Infrastructure (ADEI)**
- **Atmospheric Electric Field (AEF, Ez)**
- **Forbush Decreases (FD)**
- **Galactic Cosmic Ray (GCR)**
- **German Research Center for Geosciences (Potsdam)**
- **Geomagnetic Field (GMF)**
- **Geomagnetic Storms (GMS)**
- **Global Electric Circuit (GEC)**
- **Interplanetary Coronal Mass Ejections (ICMEs)**
- **Interplanetary Magnetic Field (IMF)**
- **The planetary Kp index estimates global geomagnetic activity levels caused by solar wind and IMF variations. Kp is compiled from measurements taken every three hours at multiple geomagnetic observatories worldwide. Each observatory assigns a local K index that is then averaged to determine the planetary Kp index.**
- **National Oceanic and Atmospheric Administration (NOAA)**
- **Near-Surface Electric Field (NSEF)**
- **Solar Cosmic Ray (SCR)**
- **Sudden Storm Commencement (SSC)**
- **The Severity of GMS Is Measured by the Kp Index and Disturbance Storm Time Index (DST)**
- **Thunderstorm Ground Enhancements (TGEs)**
- **Vertical atmospheric electric field (Ez)**